\documentclass[pra,preprint,tightenlines,showpacs,twocolumn,10pt]{revtex4} 
\usepackage{epsfig} 

\begin{document}  
    
\title{ Mutual ionization in atomic collisions 
near the electronic threshold }
 
\author{ S. F. Zhang$^1$, X. L. Zhu$^1$, A. B. Voitkiv$^{1,2}$, W. T. Feng$^1$, D. L. Guo$^1$, 
Y. Gao$^1$, R. T. Zhang$^1$, E. L Wang$^3$, and X. Ma$^1$ } 
\affiliation{ 
$^1$ Institute of Modern Physics, Chinese Academy of Sciences, 730000 Lanzhou, China \\ 
$^2$ Max-Planck-Institute for Nuclear Physics, 
Saupfercheckweg 1, D-69117 Heidelberg, Germany \\ 
$^3$  Department of Modern Physics, University of Science and Technology of China,
Hefei, 230026, Anhui, China }

\begin{abstract}

We study mutual ionization in collisions between atomic hydrogen and helium at impact velocities near the electronic threshold 
for this process (determined by the condition that 
kinetic energy of an equivelocity free electron is  
approximately equal to the sum of binding energies 
of active electrons in the projectile and target). 
We show that this process is substantially influenced 
by the Coulomb repulsion between the emitted electrons 
and that the atomic nuclei are very strongly involved  
in the momentum balance along the collision velocity.  

\end{abstract} 


\maketitle 


Two basic atomic processes 
can occur in a nonrelativistic collision 
between a bare projectile-nucleus and a target-atom. 
(i) The atom can be excited or ionized by
the impact of the projectile. 
(ii) An atomic electron can be transferred into 
a bound or low-lying continuum 
state of the projectile-ion.
The transfer can proceed with or 
without emission of radiation 
and is called radiative or  
nonradiative electron capture, respectively.

If a projectile is not fully stripped 
but carries initially an electron then in a collision 
with a target the electron can be lost \cite{mcg}-\cite{abv-buch}. 
If the loss occurs simultaneously 
with ionization of the target the corresponding process may 
be termed simultaneous projectile-target ionization. 
This process involves at least four particles 
(two nuclei and two "active" electrons) and represents 
an important case of an atomic few-body problem 
in which the interaction between the electrons 
of the projectile and target may play a crucial role 
\cite{mcg}-\cite{mut-ioniz-rel}. 

The physics of mutual ionization is simpler in fast collisions 
in which the impact velocity $v$ is much larger than the typical orbiting velocities of the projectile and target electrons. In such a case the process proceeds mainly via a single interaction between two electrons belonging to the different colliding centers (the so called two-center dielectronic interaction \cite{sdr}) whereas the nuclei (the cores) of the projectile and target are merely spectators during the collision. 
This channel of mutual ionization, which we shall call the $e$-$e$ mechanism, can be described within the first order of perturbation theory in the projectile-target interaction. 

In collisions at lower velocities another channel of mutual ionization becomes important. According to it the process proceeds via two simultaneous interactions: the electron of the target is emitted due to the interaction with the nucleus of the projectile whereas the transition of the electron of the projectile is caused by its interaction with the nucleus of the target. This reaction channel, which can be denoted as the $n$-$e$--$n$-$e$ mechanism,   
appears in a theoretical description starting with the second order perturbation expansion in the projectile-target interaction. 
In contrast to the $e$-$e$, the nuclei are strongly 
involved in the $n$-$e$--$n$-$e$ and this difference 
can be used for an experimental separation of these mechanisms 
by measuring momentum spectra of the target recoil ions 
\cite{montenegro-prl1992}-\cite{kollmus-prl2002}.  

The $e$-$e$ mechanism has an effective threshold  
corresponding to impact velocities at which 
the kinetic energy of an equivelocity free electron would 
be approximately equal to the sum of the binding 
energies of the target and projectile \cite{ee-threshold}. 
When approaching this threshold the relative contribution 
of the $e$-$e$ mechanism strongly decreases, 
below the threshold it rapidly vanishes.  

Due to very heavy nuclear masses the threshold 
for the $n$-$e$--$n$-$e$ mechanism is much lower. 
Therefore, it is widely believed 
(see e.g. \cite{sdr}, \cite{wang-2011}) that for the mutual 
ionization near the $e$-$e$ threshold 
the electron-electron interaction plays merely a minor role.    

However, in the present communication, 
where we study experimentally and theoretically 
mutual ionization in $70$ keV/u H$^0$ on He$^0$ 
collisions, it will be shown  
that the electron-electron interaction 
does substantially influence this process  
in the vicinity of the $e$-$e$ threshold. 
It will be also demonstrated 
that near the threshold  
the nuclei very actively participate 
in the momentum balance of the process  
even within the $e$-$e$ mechanism 
in which they would normally be regarded 
just as spectators. 

An impact energy of $70$ keV/u corresponds 
to a collision velocity $v=1.67$ a.u.  
This energy was chosen because a free electron 
moving with this velocity would have kinetic 
energy almost exactly equal to the sum of 
the binding energy of hydrogen and 
the first ionization potential of helium. 

In this study we shall focus on the cross section for mutual ionization differential in the longitudinal momentum of the target recoil ions  
because this cross section is not only very sensitive 
to the major aspects of the collision dynamics 
(see e.g., \cite{wu-1994}, \cite{wu-1997}-\cite{abv-2007}) 
but also yields important information in a compact form \cite{yan}. 
   
Atomic units are used throughout unless otherwise stated. 

The experiment was performed using the reaction microscope 
\cite{X1}, \cite{X2} located at the beam line of 
the $320$ keV platform for multi-discipline research with highly
charged ions \cite{X3} at the Institute of Modern Physics Lanzhou, China. Proton beams extracted from the electron cyclotron resonance ion source were accelerated to energy of $70$ keV when leaving the platform. 
Then the proton beams were collimated and transported to the collision chamber. In front of the chamber protons were
neutralized in a $20$ cm-long differentially pumped gas cell in which pure N$_2$ gas was filled with a pressure of a few $ 10^{−4} $ mbar. The protons in the beam behind the gas cell were removed by a pair of electrostatic plates downstream the cell. The deflection field was set to $2000$ V/cm 
so that the metastable H$^0$ produced in the neutralization processes could be quenched, and more than $99\%$ of them are estimated
to be in ground state \cite{15}. The remaining H$^0$ then intersected 
with the supersonic helium gas jet of the spectrometer. After the collisions, the projectiles were charge selected by another 
electrostatic deflector in front of the projectile
detector. The momenta of the target fragments were obtained via the reaction microscope. Triple coincidence measurements between 
the electron, the recoil and the projectile detectors were employed 
to distinguish the interaction channels. The momentum of the recoil ions was calibrated using two separate peaks in capture 
channel with final recoil He$^+$ ions in the ground and excited states  
and could be determined with a precision of $0.02$ a.u.

In \cite{doerner-1994}, where mutual ionization 
in $0.5$ -- $2$ MeV He$^+$ on He collisions was studied, 
it was found that one more channel noticeably contributes 
to this process at an impact energy of $500$ keV: 
emission of two electrons from He accompanied by  
capture of the electron from He$^+$. However, 
$70$ keV H$^0$ is much less effective 
in producing double ionization of He than $500$ keV He$^+$ 
and in collisions with He it is easier 
to ionize $70$ keV H$^0$ than $500$ keV He$^+$. Besides, 
the cross sections for electron capture from 
$70$ keV H$^0$ and $500$ keV He$^+$ by He$^{2+}$ are close. 
Therefore, in our case this channel 
is of minor importance and may be neglected.  

\begin{figure}[t] 
\vspace{-0.45cm}
\begin{center}
\includegraphics[width=0.55\textwidth]{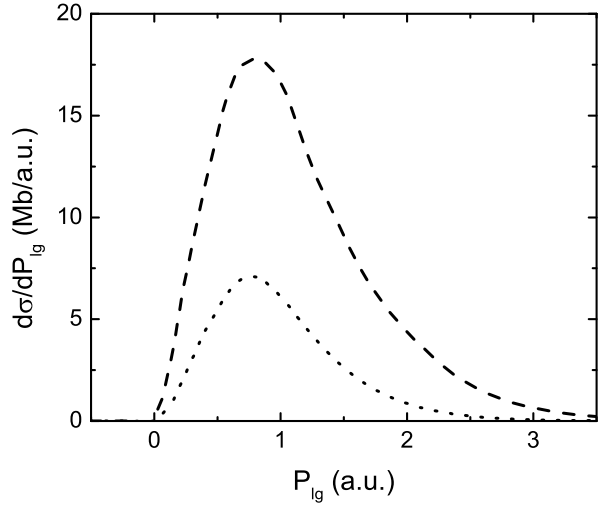}
\end{center}
\vspace{-0.5cm}  
\caption{ \footnotesize{ The cross section for mutual ionization 
differential in the longitudinal momentum of the target recoil ions  
in $70$ keV H(1s) on He(1s$^2$) collisions. 
Dot and dash curves are results of 
the first and second order calculations, respectively.}}  
\label{figure1} 
\end{figure}

In our theoretical treatment we regard the process of mutual ionization 
as an effectively four-body problem considering helium as an hydrogen-like system consisting of one (active) electron and a core with an effective charge $Z_{eff}$ determined from the first ionization potential of helium 
which results in $Z_{eff}=1.345$. The process is described using the first and second order terms of the perturbative expansion in the projectile-target interaction. In this description the initial and final states of the target (the projectile) are eigenstates of the undistorted Hamiltonian of the target (the projectile). In the second order transition amplitude 
the contributions of the first and second order terms are added  coherently.  Results obtained in this way are shown   
in figure \ref{figure1} for the cross section differential in the longitudinal momentum $P_{lg}$ of the recoil ions. In particular, 
it follows from the figure that the mutual ionization 
is dominated by the $n$-$e$--$n$-$e$ mechanism and 
that both the reaction mechanisms lead to a similar shape 
of the spectrum.     

\begin{figure}[t] 
\vspace{-0.45cm}
\begin{center}
\includegraphics[width=0.55\textwidth]{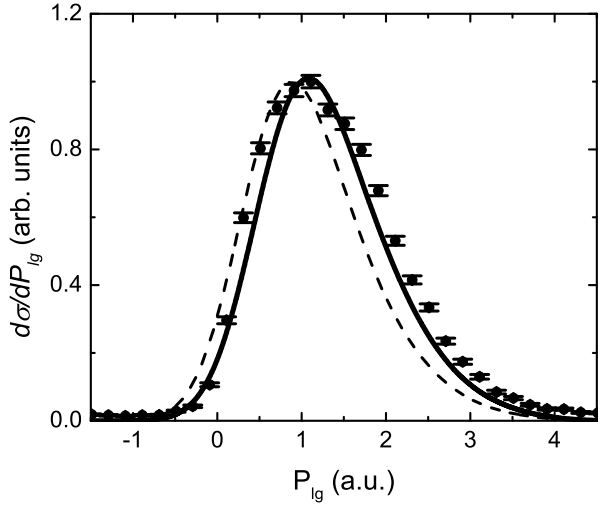}
\end{center} 
\vspace{-0.5cm} 
\caption{ \footnotesize{ The cross section for mutual ionization 
differential in the longitudinal momentum of the target recoil ions  
in $70$ keV H(1s) on He(1s$^2$) collisions. 
Symbols are the experimental results. Solid (dash) curve is   
the result of the calculation which takes into account (neglects) 
the Coulomb repulsion between the emitted electrons.}}  
\label{figure2} 
\end{figure}

In figure \ref{figure2} we compare the result of the above second order calculation (shown by a dash curve) with our experimental data. Note that 
since the experimental cross sections are not measured on absolute scale  in this figure both measured and calculated cross sections are normalized by setting their maxima to $1$. Besides, the calculated results were convoluted with an experimental resolution of $\Delta P_{lg}=0.4$ a.u..  
It is seen in the figure that the agreement between the theory and experiment is not very good: the theory overestimates the experiment at $P_{lg} \stackrel{<}{\sim} 0$ but substantially underestimates it at $P_{lg} \stackrel{>}{\sim} 1$. Besides, the positions of the maxima 
in the experimental and calculated spectra differ 
by $ \approx 0.25$ a.u. which is much larger than 
the uncertainty of $0.02$ a.u. in the experimental 
determination of this position. 

In order to understand possible reasons for this disagreement 
let us note the following. In the process of mutual ionization the initial velocities of the heavy particles -- the projectile and target nuclei -- are practically unchanged. However, the light particles -- the electrons -- do experience a substantial velocity change. The electron of the projectile, due to its interaction with the target, gets a kick opposite to the projectile motion and is, therefore, emitted from the projectile with (the longitudinal component of) a velocity noticeably smaller than that of the projectile nucleus. In turn the electron of the target in the collision gets a kick in the direction of the projectile motion and is ejected with a velocity much larger than that of the target recoil ion. 

Thus, after the collision the fastest and slowest particles are the proton and the target recoil ion (which is practically at rest), respectively, and the velocities of the electrons are in between. Important to note that when the mutual ionization occurs at impact energies close 
to the $e$-$e$ threshold the velocities of both emitted electrons are actually not very different. Moreover, since this process occurs in collisions with typical impact parameters not exceeding the size of the atoms, the electrons become close not only in the momentum but also in the position space. Therefore, one can expect that the Coulomb repulsion of the electrons in the final state may play a substantial role. 

We modelled the effect of the Coulomb repulsion by introducing 
into the (fully differential) cross section a factor $2 \pi \lambda /k_{12}/(exp(2 \pi \lambda /k_{12}) - 1)$, 
where $k_{12}$ is the relative momentum of the emitted electrons 
and $\lambda$ is a parameter. If $\lambda=1$ we obtain 
the so called Gamov factor. This factor is proportional to the absolute square of the Coulomb wave function describing the relative motion in a system of two electrons taken at zero relative distance.  

\begin{figure}[t] 
\vspace{-0.45cm}
\begin{center}
\includegraphics[width=0.55\textwidth]{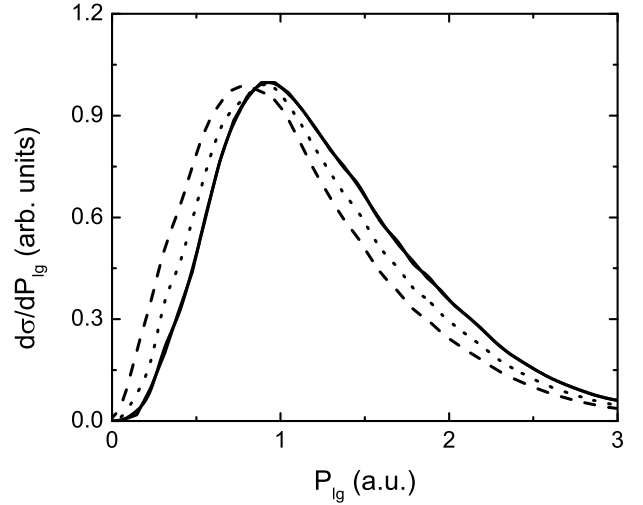}
\end{center}
\vspace{-0.5cm} 
\caption{ \footnotesize{ The cross section for mutual ionization 
differential in the longitudinal momentum of the target recoil ions  
in $70$ keV H(1s) on He(1s$^2$) collisions. 
Dash, dot and solid curves are calculated results 
obtained setting $\lambda = 0$, $0.3$ and $0.6$, respectively.}}  
\label{figure3} 
\end{figure}

From the studies of atomic ionization by electron impact 
it is known that if instead of the (full) Coulomb wave function 
only the Gamov factor is used then the repulsion effect is overestimated. 
Therefore, in our model we regarded $\lambda $ as a free parameter 
from the interval $0 \leq \lambda < 1$. Results of our calculations for the normalized cross section using three different values of $\lambda$ 
($\lambda = 0$, $0.3$ and $0.6$) are shown in figure \ref{figure3}. 

According to the calculations the Coulomb repulsion between 
the emitted electrons shifts the longitudinal momentum distribution 
of the recoil ions in the positive direction and  
also somewhat broadens it. These effects, which appear 
both in the first and second order calculations, 
could be qualitatively understood 
by noting that because of the repulsion the emitted 
electrons try to avoid each other in the momentum space. 
As a result, they tend to populate in the continuum 
a broader energy range occupying on overall higher 
energy states than it would be in the absence of the repulsion. 
Since the longitudinal momentum transfer in the collision 
is proportional to the difference between the initial 
and final energies of the electrons, the above broadening 
leads to larger longitudinal momentum transfers 
and increases their spread. Both these points are 
reflected in the shape of the longitudinal momentum 
spectrum of the recoil ions.   

The choice $\lambda = 0.6$ enables one 
to obtain the best agreement with the experimental data. 
Our calculated results with $\lambda = 0.6$, 
convoluted with the experimental resolution 
of $\Delta P_{lg} =0.4$, are shown 
in figure \ref{figure2} by a solid curve. 
It is seen that the account 
of the Coulomb repulsion brings the calculated spectrum  
in much better agreement with the experiment. 

The closeness of the electrons 
in the position and momentum space, which 
makes the Coulomb repulsion quite effective, 
can also impact the process of mutual ionization 
via the electron exchange effect. Therefore, 
we estimated the influence of 
the latter on the longitudinal momentum spectrum 
within the first order of perturbation theory. 
However, according to the estimate, the shape 
of the spectrum and the position of its maximum 
turned out to be very weakly influenced 
by the exchange effect.  

It is well known (see e.g. \cite{doerner-1994}) 
that at high impact velocities, 
where the $e$-$e$ mechanism dominates,   
the longitudinal spectrum of the recoil ions 
is almost symmetric with respect to the point $P_{lg}=0$ 
where the maximum of the spectrum is located. 
At lower velocities (but still far above the $e$-$e$ threshold) 
the $n$-$e$--$n$-$e$ channel becomes important 
that makes the spectrum asymmetric with the interval   
$P_{lg} > 0 $ being more populated compared 
to $P_{lg} < 0 $. 

An interesting peculiarity of the mutual ionization 
at impact velocities near the $e$-$e$ threshold 
is that the longitudinal spectrum of the recoil ions 
is more asymmetric, is stronger shifted to the positive 
$P_{lg}$ than at high impact velocities and that 
this holds not only for the $n$-$e$--$n$-$e$ 
but also for the $e$-$e$ reaction channel 
(see figure \ref{figure1}).    
  
The reason for this is that at impact velocities 
near the threshold the electrons, in order 
to be emitted via the $e$-$e$ mechanism, 
have to "borrow" momentum-energy from the coupling to their parenteral nuclei that affects the momentum spectra of the nuclei.    
As the collision velocity approaches the $e$-$e$ threshold, 
the internal motion of the electrons in their initial 
bound states starts to have a pronounced effect 
on the ionization process. 

Indeed, near the threshold the most favourable condition for the mutual ionization to occur is when the electron of the projectile in its initial bound state possesses a positive projection of the orbiting velocity on the direction of the projectile motion whereas the electron of the target has initially a negative component of the orbiting velocity along this motion: such a velocity configuration provides the largest relative momentum and energy for the colliding electrons. 


Since the momenta of the electron and the nucleus 
in a bound atomic state compensate each other 
(in the atomic center-of-mass frame) 
the above configuration of the electron momenta  
leads to a mirrored configuration of the momenta of the nuclei 
in which the nucleus of the projectile (target) has a negative (positive) projection of its orbiting momentum on the collision velocity in the initial bound state. Being preferential for the mutual ionization 
this configuration of the initial nuclear momenta is "selected"  
in the collisions that results, in particular, in a shift of the longitudinal momentum distribution of the target recoil 
ions towards positive $P_{lg}$.    

The peculiarities in this distribution near the $e$-$e$ threshold 
can be analyzed using the conservation of energy and longitudinal momentum in the collision that leads to the following general constraint 
on the possible values of the longitudinal momentum of 
the target recoil ion in the final state 
(no matter what the reaction channel is) 
\begin{eqnarray} 
P_{lg} \geq \frac{ k_{tr}^2/2 - \varepsilon_i}{v} + 
\frac{\epsilon_f - \epsilon_i}{v} - \frac{v}{2}.  
\label{P-rec-lim}
\end{eqnarray}
Here, $\varepsilon_i$ and $\epsilon_i$ are the initial energies 
of the electrons bound in the target and projectile,  
respectively, $\epsilon_f$ is the energy of the electron emitted 
from the projectile (as it is viewed in the rest frame of the projectile) 
and $k_{tr}$ is the absolute value of the transverse momentum 
component of the electron emitted from the target 
given in the target frame. 

At high impact velocities the constraint (\ref{P-rec-lim}) does not really set limitations but it becomes quite restrictive when the velocity approaches the $e$-$e$ threshold. For instance, for the mutual ionization in $70$ keV H on He collisions we obtain $P_{lg} \geq 0.001$ a.u. 
\cite{comment} and, thus, the spectrum of the recoil ions simply cannot have a maximum at $P_{lg} = 0 $ and be symmetric with respect to this point as it would be at high impact velocities.  

In conclusion, we have considered the process of mutual ionization in collisions of $70$ keV hydrogen with helium 
by exploring the cross section differential 
in the longitudinal momentum of the target recoil ions. 
Our results show that, although at impact velocities 
in the vicinity of the $e$-$e$ threshold 
the electron-electron interaction is not the main mechanism 
inducing mutual ionization, it nevertheless substantially influences 
this process via the Coulomb repulsion between the emitted electrons. 
Besides, we have also demonstrated that near the threshold 
the nuclei of the colliding particles are much stronger 
involved in the momentum balance of the reaction than 
it would be at high impact energies. 

{\bf Acknowledgement}.  
This work is partly supported by the Major State Basic Research Development Program of China (973 Program, Grant No. 2010CB832902) 
and by the National Nature Science Foundation of China 
under Grants Nos. 10979007 and 11274317.
A.B.V acknowledges support from the Extreme Matter Institute EMMI, 
the DFG under the project VO 1278/2-1,  
and the program for visiting international senior scientists 
of Chinese Academy of Sciences.


\begin{thebibliography}{99}
\bibitem{ee-threshold} Since the colliding electrons 
are initially bound, the threshold is broadened due 
to the Compton profiles of the corresponding bound states.
\bibitem{comment} Because of a finite resolution 
the experimental data (as well as the theory results 
after the convolution with experimental resolution) 
partly spread to $P_{lg} \stackrel{\sim}{<} 0$.
\end{thebibliography}
\end{document}